\documentclass[twocolumn,showpacs,aps,pra]{revtex4}

\newcommand{\dd}{\mathrm d}
\newcommand{\ee}{\mathrm e}

\usepackage{epsf}
\usepackage{amsmath}
\usepackage{amsfonts}
\usepackage{amssymb}
\usepackage{bm}
\usepackage{bbm}
\usepackage{nicefrac}
\usepackage{color}
\usepackage{pifont}
\usepackage{graphicx}
\usepackage{enumerate}
\usepackage{dcolumn}
\usepackage{maybemath}

\begin{document}

\newcommand{\addrMST}{
Department of Physics, Missouri University of Science and Technology,
Rolla, Missouri, MO65409-0640, USA}

\title{From First Principles of QED to an Application:\\
Hyperfine Structure of $\bm{P}$ States of Muonic Hydrogen}

\author{U.~D.~Jentschura}

\affiliation{\addrMST}

\begin{abstract}
The purpose of this article is twofold.  First, we attempt to 
give a brief overview of the different application areas of 
quantum electrodynamics (QED).
These include fundamental physics (prediction of atomic energy levels), 
where the atom may be exposed to additional external fields (hyperfine splitting and
$g$ factor). We also mention QED processes in 
highly intense laser fields, and more applied areas like Casimir and
Casimir--Polder interactions. Both the unifying aspects as well
as the differences in the the theoretical treatment required 
by these application areas (such as the treatment of infinities) 
are highlighted. Second, we discuss an application of 
the formalism in the fundamentally interesting area of the prediction 
of energy levels, namely, the hyperfine structure of $P$ states of 
muonic hydrogen.
\end{abstract}


\maketitle

%
%
\section{Introduction}

This article is divided into two parts: the first describes the intricacies and
intertwined relations of different application areas of quantum electrodynamics
(QED), and the second contains an application of the discussed concepts to a
particular problem in bound-state quantum electrodynamics, namely, the QED
corrections of relative order $\alpha(Z\alpha)^2$ to the hyperfine splitting of
$P$ states in hydrogenlike ions.

The theory of quantum electrodynamics has been developed as a general theory of
electromagnetic interactions of charged fermions and charged bosons with each
other, via the action of the quantized electromagnetic field~\cite{Di1927}.  As
such, QED has been developed as the first fully quantized theory involving both
matter fields and gauge bosons, and among all theories that comprise the
standard model, it is still the one that produces the most accurate predictions
for experiments~\cite{SaYe1990,MoTaNe2008}.

The current article is a contribution
to the special issue devoted to the 
2010 conference on the physics of
simple atomic systems (PSAS 2010), held at the Les Houches conference center
in the French Alps. Appropriately, the second part contains a 
calculation of a physical effect of relevance for one of the 
subjects discussed at the conference, namely,
the recent muonic hydrogen Lamb shift experiment at
Paul Scherrer Institute Villigen, Switzerland~\cite{PoEtAl2010}. 
In this experiment, the $2S_{1/2}\, (F=0)
\rightarrow 2P_{3/2}\, (F=2)$ hyperfine component
of the $2S \rightarrow 2P_{3/2}$ transition 
in muonic hydrogen is
measured. According to Ref.~\cite{Pa1996mu},
the transition frequency can be 
written as the sum of three contributions:
the $2S$ hyperfine structure, the
$2P_{1/2}$--$2P_{3/2}$ fine-structure splitting, and the
$2P_{3/2}$ hyperfine structure contribution.
The current paper is concerned with the 
evaluation of the QED corrections of relative order 
$\alpha (Z\alpha)^2$ to the hyperfine splitting.
The contributions to the hyperfine splitting 
in this order have recently been evaluated 
in Ref.~\cite{JeYe2010} (for $2P$ states). 
Here, we generalize the approach to $3P$ states and 
observe a smooth variation of the coefficients 
parameterizing the QED corrections with the 
principal quantum number.
Our calculation eliminates a possible 
source for an explanation of the 
observed discrepancy of experiment and theory for 
muonic hydrogen and improves the accuracy of the 
theoretical predictions for the hyperfine splitting
in muonic hydrogen.

This paper is organized as follows.
In Sec.~\ref{appl}, we 
contrast the traditional areas of quantum electrodynamics
in the description of scattering processes and in the 
calculations of energy levels in atoms with more applied 
application-oriented areas
such as highly intense laser fields
whose importance has risen in recent years.
In some sense, this section follows the personal route of the
author whose scientific 
career started in a bound-state calculation~\cite{JePa1996} and 
who added other
subfields over the course of the last ten years.
It is a brief account of both the general principles of 
quantum electrodynamics to different areas, but also, of the 
little peculiarities that have to be 
analyzed in each particular case.
This section might be criticized as containing ``only words''
and the author accepts this criticism as a possible point of view.
Then, in Sec.~\ref{hfs}, we discuss a particular 
application of the formalism, namely, the calculation 
of a QED-induced shift of the hyperfine splitting
of $P$ states in one-fermion ions. 
Muonic hydrogen is part of this class of ions.
Conclusions are reserved for Sec.~\ref{conclu}.
Throughout the article, 
we use natural units ($\hbar = c = \epsilon_0 = 1)$.

%
%
\section{From fundamental principles to applications: QED in action}
\label{appl}

QED is a versatile instrument for the description of 
processes in atoms and molecules, and also, in laser fields. 
The formalism of QED essentially relies on time-ordered perturbation
theory~\cite{MoPlSo1998}, and we may perform calculations in the theory at various 
levels of approximation. In general, a the term ``QED calculation'' 
is reserved for those electrodynamic calculations which would 
be impossible to carry out without at least quantizing one of the 
electromagnetic or fermionic fields involved in the problem.
These include, e.g., pair production processes (at tree level)
or so-called loop calculations~\cite{ItZu1980}, which describe the 
self-interactions of the quantized fields.

Indeed, the approximations which 
make a calculation possible are completely different in each
particular case and must be chosen with care.
For example, in a strong laser field, it makes no sense
to quantize the vec potential describing the laser, and classical four-vector potentials 
are the predominant means for the description of the pertinent interactions.
However, if an electron-positron  pair 
is produced in the presence of a strong laser field, then it becomes 
necessary to apply quantum theory to the particle creation and annihilation
process (i.e., to quantize the fermion field, see Ref.~\cite{LoJeKe2008}). 
By contrast, for a bound electron (in an atom) interacting with its
own radiation field (``self energy''), 
it is imperative to quantize the electromagnetic field, 
and indeed, the quantum theory of fields has found its origin and 
its first confrontation with experiment in this very subfield 
of physics~\cite{Be1947}. 

The ``laser-dressed relativistic Furry picture''
for the description of QED processes in laser fields 
has been developed over the last few years~in 
Refs.~\cite{LoJeKe2007,ScLoJeKe2007pra,LoJeKe2008,LoJeKe2009,LoJe2009prl,LoJe2009}.
In this picture, the time-dependent laser fields is
included in the unperturbed Hamiltonian that describes 
the fermions, via minimal coupling.
Formulas which enable a practical evaluation of the 
fully laser-dressed fermionic (Dirac-Volkov) propagator can be found 
in Refs.~\cite{ScLoJeKe2007pra,LoJe2009}.

The theory of quantum electrodynamics can be used both for the
description of rather ``applied'' processes with technological relevance (such
as atoms in contact with surfaces via 
Casimir--Polder interactions~\cite{Mi1994}) as well as
for the clarification of questions relevant to fundamental
physics, such as the highly accurate description of atomic energy levels in atoms.
We can attempt to 
summarize the most important application areas of QED as follows:
\begin{itemize}
\item One of the more important and traditionally established 
areas~\cite{EiGrSh2001,Ka2005}
of application concerns the spectra of bound systems (few-electron-systems, 
including excited states), as well as radiative corrections
to the electron and muon $g$ factors~\cite{HuKi1999}.
\item Also of importance are the spectra of bound systems in additional fields.
The bound electron $g$ factor~\cite{Be2000}  
is relevant for bound and free electrons under the simultaneous 
influence of the binding Coulomb field and an 
additional external magnetic field. In the calculation of the 
hyperfine splitting, one must take into account the 
Coulomb field, the nuclear magnetic field and, possibly,
the radiation field that describes the self-interaction of the 
electron.
\item We can also point to more practically applied
subfields~\cite{Mi1994,LaMNRe2006}:
Casimir interactions (between plates, atoms, or between an ionic core 
and a loosely bound Rydberg electron), and
Casimir--Polder interactions (i.e., atom-wall
interactions relevant to surface science and nanostructures).
\item Finally, QED is indispensable for the 
description of laser-related processes:
One uses a variant of the Furry picture~\cite{Fu1951}, where the (classical) laser field
is absorbed in the unperturbed Hamiltonian, 
just as much as the (classical) Coulomb field is absorbed into the 
unperturbed Hamiltonian in bound-state QED~\cite{LoJe2009}.
Laser-assisted bremsstrahlung, the channeling of electron-positron
pairs in laser fields and other fundamental processes like 
two-photon emission in intense laser fields can be described in this way.
\end{itemize}
Missing from this list is the development of effective
field theories which can sometimes leads to 
very useful approximations to QED, notably, in applications
where fields are slowly varying on the 
scale of the electron zitterbewegung frequency
$\nu_{\rm pair} = 2 m_e \, c^2/h = 2.47 \times 10^{20} \, {\rm Hz}$.
In the so-called Heisenberg-Euler Lagrangian~\cite{HeEu1936,Sc1951,DiRe1985}, 
the virtual fermion-antifermion pairs that lead to mutual
interactions of photons (with frequencies below 
$\nu_{\rm pair}$ are integrated out, and an effective 
interaction is written for the photons. 
The matching the scattering amplitudes in an 
effective and fully relativistic theory also 
is the basis for the construction of nonrelativistic QED (NRQED);
the latter theory describes bound states~\cite{CaLe1986}.
By contrast, the fully relativistic, initial formulation
of QED for bound state is described in Ref.~\cite{MoPlSo1998}.

It is surprising to see how much the above mentioned areas
can sometimes be intertwined: e.g, a numerically efficient 
code for the evaluation of Bessel functions is
required both for the calculation of spin-angular functions
in Lamb shift calculations~\cite{Ga1967,Mo1974b,JeMoSo1999}, 
but also for the numerical evaluation of the 
laser-dressed fermion propagator in intense laser fields~\cite{LoJe2009}.
The concurrence \cite{Wo1998,Ci2005}, 
which is a good measure for the
entanglement of correlated photons emitted in a two-photon
emission process in atoms~\cite{AmEtAl2009}, can also be used for the 
description of entangled photons emitted in two-photon transitions
among Dirac--Volkov states in strong laser 
fields~\cite{LoJe2009prl,LoJe2009pra}.
Another example: The multipole expansion is central to the 
expansion of the low-energy part of the bound-electron
self-energy is powers of $Z\alpha$~\cite{Pa1993,JePa1996},
where $Z$ is the nuclear charge and $\alpha$ is the 
fine-structure constant.  In Casimir-Polder atom-surface interactions, 
the multipole expansion becomes an expansion in 
powers of $a_0/z$, where $a_0$ is the Bohr radius and 
$z$ is the atom-wall distance~\cite{LaDKJe2010pra}, but 
is still based on the same principles as in bound-state
perturbation theory.

There are also important and almost trivial connections
between the calculation of QED corrections to atomic energy 
levels and QED corrections to the bound-electron $g$ factor 
and the hyperfine splitting. E.g.,
the $g$ factor can be related to an energy correction because
it relates the energy levels in an external field to the 
magnitude of the applied, external, uniform field.
The self-energy of a ``normal'' bound electron involves
one emission and one absorption process of a virtual photon.
By contrast, the additional interaction with the 
added external magnetic potential (either a uniform magnetic 
field for the $g$ factor or the nuclear dipole 
magnetic field for the hyperfine splitting)
require us to use at least third-order perturbation theory
(see also Fig.~\ref{fig1} below), with 
two interactions with the quantized radiation field,
and an additional interaction with the external field,
the latter being described on the classical level.
Consequently, related calculations,
which have progressed over a number 
of decades~\cite{Zw1961,BeEtAl2000pra,SaCh2006,SaCh2008,YeJe2010,JeYe2010,Je2010}
follow the same principles in their initial
theoretical formulation.

Nevertheless, a great deal of expertise is necessary in order 
to carry out calculations in the rather demanding subfields of 
QED. One example of a surprise is the fact that even in 
the relativistic Dirac theory of atoms, one can evaluate 
perturbations to the fully relativistic wave function
due to external magnetic fields
using generalized virial relations~\cite{Sh1991,Sh2003},
whose existence would have been hard to guess given the 
involved structure of the Dirac wave functions and of 
the Dirac--Coulomb propagator~\cite{SwDr1991a,SwDr1991b,SwDr1991c}.
One underlying theme in QED calculations
is the regularization of infinities. These infinities come
in various different forms and are not restricted to 
the famous ultraviolet divergences in quantum field 
theory which have necessitated the development of 
modern renormalization theory in the first place~\cite{ItZu1980}.
Indeed, the different origins of the infinities also 
reflect the differences among the 
various application areas of QED. Let us briefly mention
three different examples for the occurrence of infinities and 
how to deal with them within QED.
 
First example:~In the calculation of two-photon transitions of Dirac-Volkov states
(laser-assisted double bremsstrahlung~\cite{LoJeKe2007}), 
we have to regularize the intermediate
resonant Dirac-Volkov states either using their lifetime against one-photon
emission (for continuous-wave laser fields) or using a finite laser pulse duration,
calculating Dirac--Volkov states for given envelope functions of the 
laser field. The physical reason for the divergence in this case is
that there is no such thing as a perfect resonance with an
infinite lifetime.
 
Second example:~In the treatment of Casimir-Polder calculations~\cite{Mi1994}, 
we have to regularize the integrals 
that describe the interaction of an atom with the fluctuating 
electromagnetic field in the vicinity of a surface. One example is 
simply given by the integral $\int_0^\infty \dd z \, \sin(z)$.
An infinitesimal exponential convergence factor
must be introduced and leads to the regularization
\begin{equation}
\int_0^\infty \dd z \, \sin(z) \to
\lim_{\epsilon \to 0} \int_0^\infty \dd z \, \sin(z) \, \ee^{-\epsilon z} = 1 \,.
\end{equation}
The physical reason for the occurrence and regularization of the 
infinity is the obvious renormalization condition for the Casimir-Polder interaction
at infinite atom-wall distance.  

Third example:~In Lamb shift calculations~\cite{LambFirst}, we have to
regularize and renormalize the vertex calculations which describe 
the interaction of the electron with the quantized 
electromagnetic field. One possible approach is to add
counterterms to the original QED Lagrangian in order to 
ensure that the renormalized values of the mass and charge of the 
electron (after adding the radiative corrections) correspond
to the physical values. 
The physical reason for the divergence in this 
case is that is a part of
the vertex corrections is due to corrections to the electron charge (or mass)
which needs to be reabsorbed into the physical properties of the electron.

These ultraviolet divergences are not to be 
confused with the divergences in the matching parameter $\epsilon$
used in order to separate high- and low-energy parts in 
bound-state calculations~\cite{Pa1993,JePa1996}; the latter 
parameter is an asymptotic matching parameter 
(see Chap.~8 of Ref.~\cite{BeOr1978}) that is necessary 
in order to separate the regions of high-energy 
virtual photons (where the energy of the virtual photon 
is much larger than the binding energy of the electron, and an
expansion in the number of Coulomb vertices is possible)
and the regime of applicability of the multipole 
expansion (low-energy part). An illustrative example 
is given in the Appendix of Ref.~\cite{JePa2002}.

In summary, QED gives us a very powerful formalism, which can 
be applied almost universally 
to problems in atomic and laser physics, and dynamical
processes. However, the formalism needs to be adapted to each particular
application at hand, and in the course of events, more often than none, 
yet another unexpected infinity typically appears in a QED calculation,
waiting to be regularized.

\begin{figure}
\includegraphics[width=1.0\linewidth]{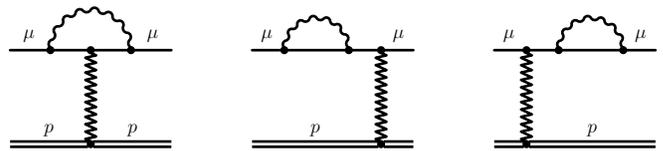}
\caption{\label{fig1} Feynman diagrams for the 
QED corrections to the hyperfine splitting in 
muonic hydrogen. The zigzag line denotes the 
exchange of a magnetic photon between muon ($\mu$)
and nucleus (proton, $p$).}
\end{figure}

%
%
\section{Self--Energy Correction to the Hyperfine Splitting of 
$\bm{P}$ States}
\label{hfs}

In view of the general formalism discussed in 
Sec.~\ref{appl}, it is perhaps useful to remember
that the hyperfine splitting is caused by the 
exchange of a magnetic photon between the atomic
nucleus and the orbiting particle.
The binding Coulomb field is taken into account 
to all orders, and the Furry picture is used.
The self energy correction to the 
hyperfine splitting involves third-order 
perturbation theory (two interactions with the 
quantized electromagnetic field and one 
interaction with the nuclear magnetic dipole field),
and the relevant Feynman diagrams are displayed in 
Fig.~\ref{fig1}.
Recently, the self-energy correction to the 
hyperfine splitting of $S$ and $P$ states
has been analyzed for bound hydrogenlike systems, in both a numerical as well as 
in an analytic approach~\cite{JeYe2006,YeJe2008,YeJe2010,JeYe2010,Je2010}.
Within the classification of QED calculations discussed in 
Sec.~\ref{appl}, the current treatment represents the case
of a traditional application to fundamental 
processes relevant for bound-state studies.

The hyperfine splitting (HFS) Hamiltonian, in the nonrelativistic 
approximation, can be written as
\begin{subequations}
\label{master}
\begin{align}
H_{\rm HFS} =& \frac{|e| m}{4 \pi} \, \vec{\mu} \cdot \vec{h} =
\frac{|e| m}{4 \pi} \,
\vec{\mu} \cdot ( \vec{h}_S + \vec{h}_D + \vec{h}_L ) \,,
\\[2ex]
\label{hS} 
\vec{h}_S =& \; 
\frac{4 \pi}{3 m^2} \, \vec{\sigma} \, \delta^3(r) \,,
\\[2ex]
\label{hD} \vec{h}_D =& \; 
\frac{ 3 \,\, \hat{r} (\vec{\sigma} \cdot \hat{r}) \, \, 
- \vec{\sigma} }{2 \, m^2 \, r^3} \,,
\\[2ex]
\label{hL}
\vec{h}_L =& \; 
\frac{\vec{L}}{m^2 \, r^3}\,.
\end{align}
\end{subequations}
Here, the vector of the $\vec\sigma$ Pauli matrices measures the 
spin of the orbiting particle, $\vec L$ measures 
the orbital angular momentum, and $\vec \mu$ is the nuclear 
magnetic dipole moment. The Fermi energy
is the nonrelativistic (NR) expectation value of an atomic state 
with quantum number $F$ for the total angular momentum 
(with projection quantum number $M_F$)
composed of the total angular momentum $j$ of the 
orbiting particle and the nuclear spin~$I$,
\begin{align}
\label{EF}
& 
E_F = \langle F M_F I j | H_{\rm hfs} | F M_F I j \rangle_{\rm NR} 
\\[2ex]
& = \alpha \, \frac{g_N}{2} \, \frac{m}{m_p} \, 
 \xi^{\rm NR}_e(j) \,
\left[ F(F+1) - I(I+1) - j(j+1) \right] \,.
\nonumber
\end{align}
Here, $m$ is the mass of the orbiting particle, and 
$m_p$ is the mass of the nucleus (proton).
The nonrelativistic value of the $\xi$ parameter 
can be inferred from Eq.~\eqref{master},
and reads for a general $nP_{1/2}$ and $nP_{3/2}$ state,
\begin{subequations}
\label{gammaNR}
\begin{align}
\label{gammaNR12}
\xi^{\rm NR}_e(\tfrac12) \equiv & \;
\xi^{\rm NR}_e(nP_{1/2}) =
\frac{4}{9} \, \frac{(Z\alpha)^3 m}{n^3}\,,
\\
\label{gammaNR32}
\xi^{\rm NR}_e(\tfrac32) \equiv & \;
\xi^{\rm NR}_e(nP_{3/2}) =
\frac{4}{45} \, \frac{(Z\alpha)^3 m}{n^3}\,.
\end{align}
\end{subequations}
The QED self-energy
correction to the HFS can be conveniently expressed in terms of
multiplicative corrections to the quantity $\xi^{\rm NR}_e(j)$,
\begin{equation}
\label{defgamma}
\xi^{\rm NR}_{\rm e}(j) \to
\xi^{\rm NR}_{\rm e}(j) \,
\left[ 1 + \delta \xi_e(j) \right] \,.
\end{equation}
The corresponding corrections to the position of the HFS sublevels are
\begin{equation}
\label{defEHFS}
\delta E_{\rm HFS} = E_F\, \delta \xi_e(j) \,,
\end{equation}
As explained in Ref.~\cite{JeYe2010}, the 
QED self-energy correction to the HFS can be 
decomposed into a high- and a low-energy part.
The high-energy part can be treated using
electron form factors and reads, for general
$nP$ states,
\begin{subequations} 
\begin{align} 
\label{genHP12}
\delta \xi^{\rm H}_e(\tfrac12) =& \;
\frac{\alpha}{\pi} \Biggl\{
\frac14 + (Z\alpha)^2 \left[ \frac{19}{144} + \frac{7}{8 \, n} +
\frac{5}{16 n^2} \right.
\nonumber\\
& \; \left. - 2 \frac{n^2-1}{n^2} \ln\left( \frac{m}{2 \epsilon} \right)
\right]  \Biggr\}\,,
\\[2ex]
\label{genHP32}
\delta \xi^{\rm H}_e(\tfrac32) =& \;
\frac{\alpha}{\pi} \Biggl\{
-\frac18 + (Z\alpha)^2 \left[ -\frac{109}{2880} - \frac{11}{32 \, n} +
\frac{1}{30 n^2} \right] \Biggr\} \, .
\end{align}
\end{subequations} 
For $n=3$, we obtain
\begin{subequations}
\begin{align}
\delta \xi^{\rm H}_e(3P_{1/2}) =& \;
\frac{\alpha}{\pi} \Biggl\{
\frac14 + (Z\alpha)^2 \left( 
- \frac{16}{9} \, \ln\left( \frac{m}{2 \, \epsilon} \right) 
+ \frac{11}{24} \right) \Biggr\} \,.
\end{align}
The low-energy part 
can be expressed as a correction to the Bethe logarithm~\cite{Be1947}
due to the HFS Hamiltonian~\eqref{master}.
As compared to the calculation of the unperturbed 
Bethe logarithm~\cite{JeMo2005bethe},
we have to introduce a perturbative correction to the Schr\"{o}dinger 
binding energy $E$ due to the HFS interaction,
and a further correction due to the 
Hamiltonian $H$ which is perturbed by the HFS.
Furthermore, the nonrelativistic bound-state 
wave function $\Psi$ also receives a correction due to the 
HFS Hamiltonian.
For $n=3$, the individual results for the contributions to the 
low-energy part are found here and read
\begin{align}
\delta \xi^{\rm L}_E(3P_{1/2}) =& \;
\frac{\alpha}{\pi} (Z\alpha)^2 
\left\{ - \frac{2}{27} \, \ln\left( \frac{\epsilon}{(Z\alpha)^2 \, m} \right) 
- 0.10840 \right\} \,,
\\[2ex]
\delta \xi^{\rm L}_H(3P_{1/2}) =& \;
\frac{\alpha}{\pi} (Z\alpha)^2 
\left\{ -\frac{10}{27}  \, \ln\left( \frac{\epsilon}{(Z\alpha)^2 \, m} \right) 
+ 0.44038  \right\} \,,
\\[2ex]
\delta \xi^{\rm L}_\Psi(3P_{1/2}) =& \;
\frac{\alpha}{\pi} (Z\alpha)^2 
\left\{ -\frac43  \, \ln\left( \frac{\epsilon}{(Z\alpha)^2 \, m} \right) 
+ 2.37607 \right\} \,,
\\[2ex]
\delta \xi^{\rm L}_J(3P_{1/2}) =& \;
\frac{\alpha}{\pi} (Z\alpha)^2 
\left( -0.58478 \right) \,.
\end{align}
The total result for the low-energy part thus is 
\begin{align}
\delta \xi^{\rm L}_e(3P_{1/2}) =& \;
\frac{\alpha}{\pi} (Z\alpha)^2 
\left\{ - \frac{16}{9} \, \ln\left( \frac{\epsilon}{(Z\alpha)^2 \, m} \right) 
+ 2.12328 \right\} \,.
\end{align}
\end{subequations}
Adding the high- and low-energy parts, the 
overlapping parameter $\epsilon$ (see Sec.~\ref{appl}) 
cancels, and we obtain the self-energy correction
\begin{align}
& \delta \xi_{\rm SE}(3P_{1/2}) =
\delta \xi^{\rm H}_e(3P_{1/2}) +
\delta \xi^{\rm L}_e(3P_{1/2}) 
\nonumber\\[2ex]
& \; = \frac{\alpha}{\pi} 
\left\{ \frac14 + (Z\alpha)^2 
\left[ - \frac{16}{9} \, \ln\left[ (Z\alpha)^{-2} \right]
+ 3.81388 \right] \right\} \,.
\end{align}
Together with a numerically small vacuum polarization correction,
which also has been evaluated in Ref.~\cite{JeYe2010}, 
\begin{equation}
\delta \xi_{\rm VP}(3P_{1/2}) =
\frac{\alpha}{\pi} (Z\alpha)^2 
\left( \frac25 \, \frac{n^2 - 1}{n^2} \right)  \,,
\end{equation}
we evaluate the QED correction of relative 
order $\alpha (Z\alpha)^2$ as
\begin{align}
& \delta \xi_e(3P_{1/2}) =
\delta \xi_{\rm SE}(3P_{1/2}) +
\delta \xi_{\rm VP}(3P_{1/2}) 
\nonumber\\[2ex]
& \; = \frac{\alpha}{\pi} \left\{ \frac14 + (Z\alpha)^2 
\left[ - \frac{16}{9} \, \ln\left[ (Z\alpha)^{-2} \right]
+ 4.16943 \right] \right\} \,.
\end{align}
According to Eq.~\eqref{genHP32},
the high-energy part for the $3P_{3/2}$ state is 
free from any logarithmic terms and reads
\begin{subequations}
\begin{align}
\delta \xi^{\rm H}_e(3P_{3/2}) =& \;
\frac{\alpha}{\pi} \Biggl\{
-\frac18 + (Z\alpha)^2 \left( -\frac{257}{1728} \right)
\Biggr\} \,.
\end{align}
For $3P_{3/2}$, the individual results for the contributions to the
low-energy part are given as 
\begin{align}
\delta \xi^{\rm L}_E(3P_{3/2}) =& \;
\frac{\alpha}{\pi} (Z\alpha)^2 
\left\{ - \frac{2}{27} \, \ln\left( \frac{\epsilon \, m^{-1}}{(Z\alpha)^2} \right) 
- 0.10840 \right\} \,,
\\[2ex]
\delta \xi^{\rm L}_H(3P_{3/2}) =& \;
\frac{\alpha}{\pi} (Z\alpha)^2 
\left\{ \frac{38}{27}  \, \ln\left( \frac{\epsilon}{(Z\alpha)^2 \, m} \right) 
- 0.34371 \right\} \,,
\\[2ex]
\delta \xi^{\rm L}_\Psi(3P_{3/2}) =& \;
\frac{\alpha}{\pi} (Z\alpha)^2 
\left\{ -\frac43  \, \ln\left( \frac{\epsilon}{(Z\alpha)^2 \, m} \right) 
+ 2.37607 \right\} \,,
\\[2ex]
\delta \xi^{\rm L}_J(3P_{3/2}) =& \;
\frac{\alpha}{\pi} (Z\alpha)^2 
\left( -1.46194 \right) \,.
\end{align}
The total result for the low-energy part is
\begin{align}
\delta \xi^{\rm L}_e(3P_{3/2}) =& \;
\frac{\alpha}{\pi} (Z\alpha)^2 
\left( 0.46202 \right) \,,
\end{align}
\end{subequations}
One has to perform the calculation carefully because the 
cancellation of the logarithmic term, which normally serves as 
a check in analytic evaluation of QED effects in atoms, 
cannot be used here as the logarithm accidentally cancels.
Adding the high- and low-energy parts, we finally obtain
for the $3P_{3/2}$ state,
\begin{align}
\delta \xi_{\rm SE}(3P_{3/2}) =& \;
\delta \xi^{\rm H}_e(3P_{3/2}) +
\delta \xi^{\rm L}_e(3P_{3/2}) 
\nonumber\\[2ex]
=& \; \frac{\alpha}{\pi} 
\left\{ -\frac18 + (Z\alpha)^2 
\left( 0.31329 \right) \right\} \,.
\end{align}
The vacuum polarization correction of order $\alpha (Z\alpha)^2$ 
vanishes, as detailed in Ref.~\cite{JeYe2010},
\begin{equation}
\delta \xi_{\rm VP}(3P_{3/2}) = 0 \,,
\end{equation}
and we have for the 
total QED effect of relative order $\alpha (Z\alpha)^2$,
\begin{equation}
\delta \xi_e(3P_{3/2}) =
\frac{\alpha}{\pi} \left\{ -\frac18 + (Z\alpha)^2 \, \left( 0.31329 \right) \right\} \,.
\end{equation}
It is instructive to compare the results
\begin{align}
\label{R1}
\delta \xi_e(3P_{1/2}) =& 
\frac{\alpha}{\pi} \left\{ \frac14 
+ (Z\alpha)^2 
\left[ - \frac{16}{9} \, \ln (Z\alpha)^{-2} 
+ 4.16943 \right] \right\} ,
\nonumber\\[2ex]
\delta \xi_e(3P_{3/2}) =& 
\frac{\alpha}{\pi} \left\{ -\frac18 + (Z\alpha)^2 \, \left( 0.31329 \right) \right\} \,,
\end{align}
with the corresponding result for $2P$ states from 
Ref.~\cite{JeYe2010},
\begin{align}
\label{R2}
\delta \xi_e(2P_{1/2}) =& 
\frac{\alpha}{\pi} \left\{ \frac14 
+ (Z\alpha)^2 
\left[ - \frac32 \, \ln (Z\alpha)^{-2} 
+ 3.70343 \right] \right\} ,
\nonumber\\[2ex]
\delta \xi_e(2P_{3/2}) =& 
\frac{\alpha}{\pi} \left\{ -\frac18 + (Z\alpha)^2 \, \left( 0.17198\right) \right\} \,.
\end{align}
As usual in QED calculations, the magnitude of the coefficients
grows with the principal quantum number.
For the hyperfine splitting of $P$ states in muonic hydrogen,
the corrections of order $\alpha (Z\alpha)^2$ 
listed in Eqs.~\eqref{R1} and~\eqref{R2} 
entail only a $10^{-6}$ correction and are thus only of marginal
significance. However, they imply that a 
possibly large double logarithmic correction, whose existence had been 
conjectured in Refs.~\cite{SaCh2006,SaCh2008}, actually vanishes and cannot
possibly contribute to the explanation of the observed experimental-theoretical
discrepancy observed in Ref.~\cite{PoEtAl2010}.

%
%
\section{Conclusions}
\label{conclu}

In Sec.~\ref{appl}, we have tried to 
elucidate general aspects of QED theory, 
and also, to illustrate the application of these
concepts to a particular effect of relevance for the 
spectrum of muonic hydrogen. We have stressed the 
versatility of quantum electrodynamics for the 
description of processes in 
atomic and laser physics.
Within the different application areas in
bound-state theory, Casimir interactions,
and laser-related processes, there are underlying
and unifying concepts, but each area is characterized by 
different suitable approximations. 
The Furry picture has recently been generalized and applied
to the case of a strong dressing laser field
within a fully relativistic formalism.
Indeed, the laser-dressed Furry picture
represents the appropriate formalism for cases
when matter perturbs light (and not the other way around).
In the bound-state Furry picture,
the binding Coulomb field plays a dominant 
role and is included into the unperturbed Hamiltonian 
of the electron-positron (fermion) field.
The fundamental field operators describing the 
fermion fields are constructed using 
solutions of a Dirac equation that includes 
the nonperturbative, classical, background field
(laser or Coulomb). Still, there are huge differences in the 
treatment of various sources of divergences and infinities
that occur in the different subfields of QED.
The different physical reasons for their occurrence
are also discussed in Sec.~\ref{appl}.

In Sec.~\ref{hfs}, we apply the Furry picture
formalism to the calculation of the 
QED (self energy and vacuum polarization) 
correction of relative order $\alpha (Z\alpha)^2$
to the hyperfine splitting of $P$ states
in hydrogenlike systems. Our results 
given in Eq.~\eqref{R1} for $3P_{1/2}$ and $3P_{3/2}$ states
confirm a general trend: the magnitude of the 
coefficients grows with the principal quantum number,
and the logarithmic terms have been found to be 
in agreement with the general formulas 
derived in Ref.~\cite{JeYe2010} for general 
principal quantum numbers.
The final result eliminates a possible source for a 
shift in the theoretical predictions in view of a 
conjectured double logarithmic term~\cite{SaCh2006,SaCh2008},
which however is shown to be absent in the 
relative order $\alpha (Z\alpha)^2$.
The corrections of relative order $\alpha$ have already been 
included into the theoretical predictions for the 
hyperfine splitting in Ref.~\cite{Pa1996mu} and are confirmed here.

Finally, let us remark that in view of the discrepancy 
between theory and experiment recently observed 
in muonic hydrogen, it would be very interesting to 
proceed in the high-precision QED experiments on 
muonic systems, and, in particular, to
realize the long planned experiment on dimuonium, 
or true muonium~\cite{BrLe2009}, which has 
been analyzed theoretically in a number of 
publications~\cite{Ma1987,JeSoIvKa1997,KaJeIvSo1998}.

\section*{Acknowledgments}

The author acknowledges helpful conversations with 
V. A. Yerokhin and K. Pachucki. 
Support from the National Science Foundation 
and from the National Institute of Standards and Technology
(Precision Measurement Grant) is gratefully acknowledged.

\end{document}